\documentclass[prl,reprint,showkeys,showpacs]{revtex4-1}

\usepackage{graphicx}
\usepackage{dcolumn}
\usepackage{bm}
\usepackage{amssymb}
\usepackage{amsmath}
\usepackage{pstricks}

\def\={\;\;=\;\;}                               

\renewcommand{\mathbf}[1]{\boldsymbol{#1}}

\def\wt#1{\widetilde{#1}}                       
\def\({\left(}                                  
\def\){\right)}                                 
\def\[{\left[}                                  
\def\]{\right]}                                 

\def\wh#1{\widehat{#1}}

\def\wh#1{\widehat{#1}}

\def\rr{{\mathbf{r}} }
\def\III{{\cal I}}
\def\SSS{{ \cal S}}
\def\CC{{\cal C}}
\def\II{{\cal J}}
\def\MM{{\cal M}}
\def\n{{n_1}}
\def\nn{{n_2}}
\def\NN{{\cal N}}
\def\AA{{\cal A}}
\def\tt{s}
\def\sss{\tau }
\def\fz{{ \cal Q }_\nu}
\def\TE{\mbox{\tiny TE}}
\def\TM{\mbox{\tiny TM}}
\def\ccc{\widehat{{\cal C}}}

\begin{document}
\title{Numerical Regularization of Electromagnetic Quantum Fluctuations in Inhomogeneous Dielectric Media}
\author{Shin-itiro Goto}
\author{Alison C. Hale}
\author{Robin W. Tucker}
\author{Timothy J. Walton}
\affiliation{Department of Physics, University of Lancaster and Cockcroft Institute, Daresbury Laboratory, Warrington, UK }
\date{\today}
\begin{abstract}
Electromagnetic Casimir stresses are of relevance to many technologies based on mesoscopic devices such as MEMS embedded in dielectric media, Casimir induced friction in nano-machinery, micro-fluidics and molecular electronics. Computation of such stresses based on cavity QED  generally require numerical analysis based on a regularization process. A new scheme is described that has the potential for wide applicability to systems involving realistic inhomogeneous media. From a knowledge of the spectrum of the stationary modes of the electromagnetic field the scheme is illustrated by estimating numerically the Casimir stress on opposite faces of a pair of perfectly conducting planes separated by a vacuum and the change in this result when the region between the plates is filled with an incompressible {\it inhomogeneous} non-dispersive dielectric.
\end{abstract}
\maketitle

\section{Introduction}
Electromagnetic interactions between electrically neutral isolated polarizable sources such as atoms or molecules are often referred to as Van der Waals forces. In a quantum field theoretical description they give rise to Casimir forces, particularly when some of the sources are replaced by a neutral continuum. Such a continuum may be restricted to interfaces between different regions of space and some regions may contain polarizable media with conducting or dispersive properties. In such cases one is confronted with the problem of calculating quantum induced stresses in such regions and the resulting pressures on the surrounding interfaces \cite{Milton1,Milton2}. In situations where the sources are idealized to occupy perfectly conducting surfaces, it is possible to estimate the induced Casimir stresses on the surfaces in terms of quantum fluctuations of the electromagnetic field in the vacuum \cite{Bordag}. For a pair of approximately parallel conducting planes such integrated stresses have been detected experimentally and their dependence on the separation between the planes measured. However, the analytic derivation of vacuum induced Casimir stresses on conducting surfaces with non-planar geometries is notoriously difficult  \cite{Boyer} to ascertain with confidence. These difficulties are compounded in situations where such surfaces bound dielectric media that may be dispersive  \cite{Buhmann,Philbin1} or contain inhomogeneous \cite{Leonhardt1,Leonhardt2} magnetic and electric susceptibilities \cite{Inui1,GTW}. Furthermore, the conceptual basis on which such calculations are expected to be reliable depends on whether it is reasonable to treat polarizable systems as a continuum when quantum effects become significant. However, one expects a continuum model of rigid dielectric media to be reasonable in mesoscopic systems where bounding geometries cannot resolve molecular detail. In such cases, the response of the medium to electromagnetic fluctuations is given in terms of piecewise smooth susceptibility tensors with components that may depend non-linearly and non-locally on space and time. For media with linear piecewise homogeneous non-conducting response functions, Lifshitz \cite{Lifshitz2,Lifshitz3,Kampen} developed a phenomenological scheme based on the analytic properties of Green tensors. Since its inception, this theory has not been significantly refined to deal with more general physical systems that are now of relevance in a number of modern technologies. These include the influence of Casimir stresses due to complicated geometries in MEMS devices embedded in inhomogeneous or non-linear dielectric media, Casimir induced friction \cite{Pendry} in nano-machinery, micro-fluidic and molecular electronic devices. Perhaps the most significant feature of the Lifshitz theory limiting its applicability to such systems is its reliance on a detailed knowledge of a Green tensor (and its analytic structure) leading to a viable regularization scheme. All attempts to apply the methods of cavity QED to mesoscopic systems containing dielectrics with inhomogeneous permittivities also rely on a knowledge of the quantum Hamiltonian of the electromagnetic field in the medium in order to calculate finite quantum expectation values of ``observables'' that can be compared with experiment. Since any quantum field is an infinite dimensional dynamical system, such values need to be determined by a regularization process that discards unobservable self-forces between sources \cite{Visser}. When canonical dynamical variables can be chosen so that the Hamiltonian for the electromagnetic field has a discrete angular frequency spectrum $\{ \omega_\rr \}$ and the same structure as the Hamiltonian describing an infinite number of simple harmonic oscillators at each point in the medium, the regularization of the (zero temperature) ground state energy  $ \frac{1}{2}\hbar\sum_{\rr} \omega_{\rr}$ is often {\it defined} by continuing to $s=-1$ the function $\frac{1}{2}\hbar\zeta(s)$ where
\begin{eqnarray*}
    \zeta(s) &=& \sum_{\rr} \omega_{\rr}^{-s} = \frac{1}{\Gamma(s)} \int_{0}^{\infty} \,dt\, t^{s-1}\sum_{\rr} \, e^{ -\omega_{\rr}t }
\end{eqnarray*}
with $s>s_{0}$ for some $s_{0}$ that renders the integral convergent. However, there are few cases in which the sum $\sum_{\rr} \, e^{ -\omega_{\rr}t }$ can be performed analytically and so the analytic continuation becomes difficult and recourse to numerics is often inevitable \cite{Wolfram}. Alternative regularization schemes \cite{Reuter} bypass the ground state energy and employ point-splitting techniques on components of the electromagnetic stress-tensor prior to numerical analysis. Such approaches are not universally applicable \cite{Milton1} and where used often employ the truncation of an infinite series whose radius and rate of convergence is rarely known.

\section{Laurent Regularization}
It is the aim of this letter to offer a more robust numerical algorithm that has the potential for wider applicability than existing regularization schemes. It will be illustrated by estimating numerically the original Casimir stress on opposite faces of a pair of perfectly conducting planes separated by a vacuum and the change in this result when the region between the plates is filled with an inhomogeneous non-dispersive dielectric. It requires for its implementation a knowledge of the spectrum of the stationary modes of the electromagnetic field between the plates. Such a spectrum will in general arise as the infinite number of roots of a system of transcendental equations obtained from imposing the appropriate boundary conditions for the modes of the electromagnetic field. This involves solving a classical boundary value problem for stationary states. For a simply connected bounded rigid rectangular cavity with perfectly conducting walls such modes can be classified as either TE or TM with, in general, different spectra. They are used to construct a Fock space of electromagnetic modes in a gauge in which the quantum Hamiltonian is quadratic in creation and annihilation operators for each mode. For a stationary system there is no ambiguity in the classical Maxwell stress tensor for the electromagnetic field in a dielectric, so one can write down the ground state expectation value of any component of this tensor in the medium. Evaluated on one side of any plane surface this yields, after regularization, the quantum stress on that side. In an inhomogeneous dielectric, such stresses will in general be different on different faces of the cavity. The simplest realistic situation is to suppose that the permittivity varies smoothly in only one direction in a rigid 3-dimensional rectangular box with perfectly conducting boundaries. If the dimensions of the box are such that two opposite pairs of end planes have much smaller areas than the pair separated by the direction of the inhomogeneity it can be shown that the difference in end pressures in this direction requires the regularization of a couple of double integrals, each of the form $\int_{0}^{\infty}\int_{0}^{\infty} {\cal F}(x,y)\,dx\,dy $. Each ${\cal F}(x,y)$ is determined by a spectrum generating function chosen so that each integral
\begin{eqnarray*}
    \II(s) &=& \int_{y_{0}}^{\infty}\int_{x_{0}}^{\infty} \, e^{-s(x+y)  }{\cal F}(x,y)\,dx\,dy
\end{eqnarray*}
regarded as a function in the complex $s$-plane is analytic in an annular region centered on $s=0$. As such, it admits a representation as a Laurent expansion in this domain. The regularized value of $\II(0)$ is then defined to be the term $c_{0}$ in this Laurent expansion that is independent of $s$ and corresponds to discarding the principal part of the Laurent series before taking the limit as $s$ tends to zero. The problem is how to determine numerically $c_{0}$ from a numerical computation of $\II(s)$ when neither the principle part of its Laurent expansion about $s=0$ nor its domain of convergence is known a priori. When $\II(s)$ is meromorphic with a pole of order $N_{1}$ at $s=0$
\begin{eqnarray}\label{LSERIES}
    \II(s) &=& \sum_{n=N_{1}}^{\infty} \, c_{n} s^{n}
\end{eqnarray}
for real $s\in [\epsilon_{s}, s_{R}]$, finite negative integer $N_{1}$ and constants $0<\epsilon_{s}\ll 1, \, s_{R}>0$.
The algorithm for estimating $c_{0}$ proceeds first by discretizing the range $s\in [\epsilon_{s}, s_{R}]$  for some $s_{R}$  to generate the set $\SSS=\{ s_{j}\, \vert\, 1<j<J\}$ and then evaluating $\II(s)$ numerically at $\SSS$ to generate the set $\III=\{ \II_{j}=\II(s_{j}) \,\vert\, 1<j<J\}$. Let $\AA$ denote a matrix where each element represents a truncated Laurent series of the form:
\begin{eqnarray}
    L_{n_{1}}^{n_{2}}(s) &=& \sum_{n=n_{1}} ^{n_{2}}\, c_{n}(n_{1},n_{2}) s^{n}   \qquad\qquad  n_{1} < n_{2}
\end{eqnarray}
for some positive integer $n_{2}$ and negative integer $n_{1}$. To effect a numerical fit of $\II(s)$ to the ``appropriate'' Laurent series, one first determines the matrix elements in any sub-matrix of $\AA$ by fitting the data $\III,\SSS$ to each $L_{n_{1}}^{n_{2}}(s) $ by linear regression. Thus, for
all integer ranges from $n_{1}$ to $n_{2}$ with $N_{1} < n_{1} \le -1$ and $1 \le n_{2} < N_{2}$ one may calculate the set $\{c_{n}(n_{1},n_{2})\}$ associated with each matrix element. We seek criteria such that $c_{0}$ in (\ref{LSERIES}) is approximated by some $c_{0}(n_{1},n_{2})$ in $\AA$, for an ``optimal'' choice of the integers $N_{1},N_{2}$ defining the size of the sub-matrix. The strategy is then to prune the principal part of each matrix element by filtering out of each truncated Laurent series those terms with coefficients $c_{n}(n_{1},n_{2})$ with $n<0$ that satisfy
\begin{eqnarray*}
    \frac{ \vert c_{n}(n_{1},n_{2})  \vert }{\MM_{n}(n_{1},n_{2})  } < \epsilon_{c}
\end{eqnarray*}
for some tolerance $0<\epsilon_{c}\ll 1$ where the average $\MM_{n}(n_{1},n_{2}) = \frac{1}{\vert n\vert }\sum _{j=n_{1}}^{j=-1} \,c_{n}(j,n_{2})$. After this filtering process, one has a sequence of truncated Laurent series $\{ \wh{L}_{n_{1}}^{n_{2}}(s) \}$ whose principal parts contain only the coefficients
\begin{eqnarray*}
    \CC_{\n}^{\nn} &=& \left\{ c_{n}(\n,\nn) \, \left\vert \, \frac{ \vert c_{n}(n_{1},n_{2}) \vert }  {\MM_{n}(n_{1},n_{2}) } > \epsilon_c \right.\right\}.
\end{eqnarray*}
Let $C_{\NN}(\n,\nn)$ denote the coefficient of the most singular term in each $\{ \wh{L}_{n_{1}}^{n_{2}}(s)\}$ where by hypothesis $\NN<0$. By examination of this coefficient for all $N_{1} \le n_{1} \le -1$ and $1\le n_{2} \le N_{2}$ in the selected sub-matrix, one may discover a new sub-matrix whose elements contain truncated Laurent series with the same fixed value of $\NN$. Next, for each positive integer $\nn$, define the data set
\begin{eqnarray*}
    \wh{\III}(\nn) &=& \{ \wh{\II}_{j}(n_2)\, \vert \, 1<j <J  \}
\end{eqnarray*}
where $\wh{\II}_{j}(\nn)=\II_{j} - C_{\NN}(\NN, \nn)s_{j}^{\NN}$ and use linear regression again to fit this to a sequence of new truncated Laurent series
\begin{eqnarray*}
    {\cal{L}}_{\NN}^{\wh{n}_2  }(s,n_{2}) &=& \sum_{n=\NN}^{\wh{n}_{2}} \wh{\cal{C}}_{n}(\NN, \wh{n}_{2},n_{2}) \, s^{n}
\end{eqnarray*}
with $1 \le \wh{n}_{2}, n_2 \le N_{2}$. If, for such a fixed $\NN$, one connects the points obtained by plotting $\wh{\cal{C}}_{0}(\NN, \wh{n}_{2},n_{2})$ against $\wh{n}_{2}$ for each $n_{2}$ in the range $1\le n_{2} \le N_{2}$, curves are generated that attain turning points (or asymptotes) with ordinates at $ \wt{c}_{0}(\NN,\wh{n}_{2}, n_{2})$ in close proximity. As a result, an estimate of the required Casimir coefficient $c_{0}(\NN)$ is defined to be $\frac{1} {N_{2}}\sum_{n_{2}=1}^{N_2}\wt{c}_{0}(\NN,\wh{n}_{2},n_{2})$.

\section{Applications}
If $F(z)$ is meromorphic, $f(z)$ entire and well behaved in the complex $z$-plane where $0\leq\arg(z)\leq{\pi}/{2}$ and $\{z_{r}\}$ denotes simple roots of the equation $F(z)=0$, then by Cauchy's theorem
\begin{eqnarray*}
    \sum_{r} f(z_r) &=& -\frac{1}{\pi} \int_{0}^{\infty} f(i y)\Delta(iy) \,dy
\end{eqnarray*}
where
\begin{eqnarray*}
    \Delta(z) &=& \frac{F^{\prime}(z,\ldots)}{F(z,\ldots)}.
\end{eqnarray*}
For the empty rectangular box with sides of length $L_{x},L_{y},L_{z}$, the TE and TM spectra in the box are degenerate with:
\begin{eqnarray*}
    \frac{\omega_{\mathbf n}^{2}}{c^{2}} &=& k_{x}^{2} + k_{y}^{2} + \frac{n_{z}^{2}\pi^{2}}{L_{z}^{2}} \quad \text{where} \quad k_{x}=\frac{n_{x}\pi}{L_{x}}, \quad k_{y}=\frac{n_{y}\pi}{L_{y}},
\end{eqnarray*}
and $n_{x},n_{y},n_{z} \in Z^{+}$. With  $k_{x}^{2}+k_{y}^{2}=k^{2}$, a suitable spectrum generating function is
\begin{eqnarray*}
    F(\omega, k ) &=& \sin\left( L_{z} \sqrt{\left( \frac{\omega^{2}}{c^{2}} - k^{2}\right)}\right).
\end{eqnarray*}
Then, for $L_{x}, L_{y} \gg L_{z}$ (i.e. parallel plates) and $f(z)={\hbar z}/{2}$,
the TE Casimir energy $\langle{\cal E}^{\TE}\rangle$ is determined from a Laurent expansion of
\begin{eqnarray*}
    E^{\TE}(s) &=& -\frac{L_{x}L_{y}\hbar c}{4\pi^{2} L_{z}^{3}}\, \,{\cal I}(s)
\end{eqnarray*}
where
\begin{eqnarray}\label{data1}
    {\cal I}(s) &=& \frac{1}{3}\int_{0}^{\infty}\, r^{3} e^{-s r} \coth(r)\,dr.
\end{eqnarray}
For $s>0$, the integral (\ref{data1}) can be performed analytically and for small $s$ determines all coefficients in the Laurent series
\begin{eqnarray*}
    {\cal I}(s) &=& \sum_{n=-4}^{\infty} c_{n} s^{n}.
\end{eqnarray*}
The TE Casimir energy is then taken to be
\begin{eqnarray*}
    \langle{\cal E}^{\TE}\rangle &=& -\frac{L_{x} L_{y} \hbar c}{4\pi^{2}L_z^{3}} c_{0}.
\end{eqnarray*}
In this case, the value of $c_{0}$ can be determined analytically since the integral can be written in terms of the third derivative of the polygamma function $\Psi(3,x)$ as \footnote{$\Psi(3,x) = \frac{\partial^{4}\ln( \Gamma(x)) }{\partial x^{4}}$}
\begin{eqnarray*}
    {\cal I}(s) &=& \frac{1}{24}\Psi\left(3,\frac{s}{2 }\right) - \frac{2}{s^{4}}.
\end{eqnarray*}
The Laurent expansion of this expression gives the exact value $c_{0}={\pi^{4}}/{360}$. If the integral (\ref{data1}) is evaluated numerically to 7 significant figures for a range of $s$ in the vicinity of the origin, one obtains from our algorithm good agreement with this analytic result. It agrees with the value determined by Riemann $\zeta$ function regularization within 1.2\%.
\begin{figure}[h]
    \includegraphics[width=0.8\columnwidth]{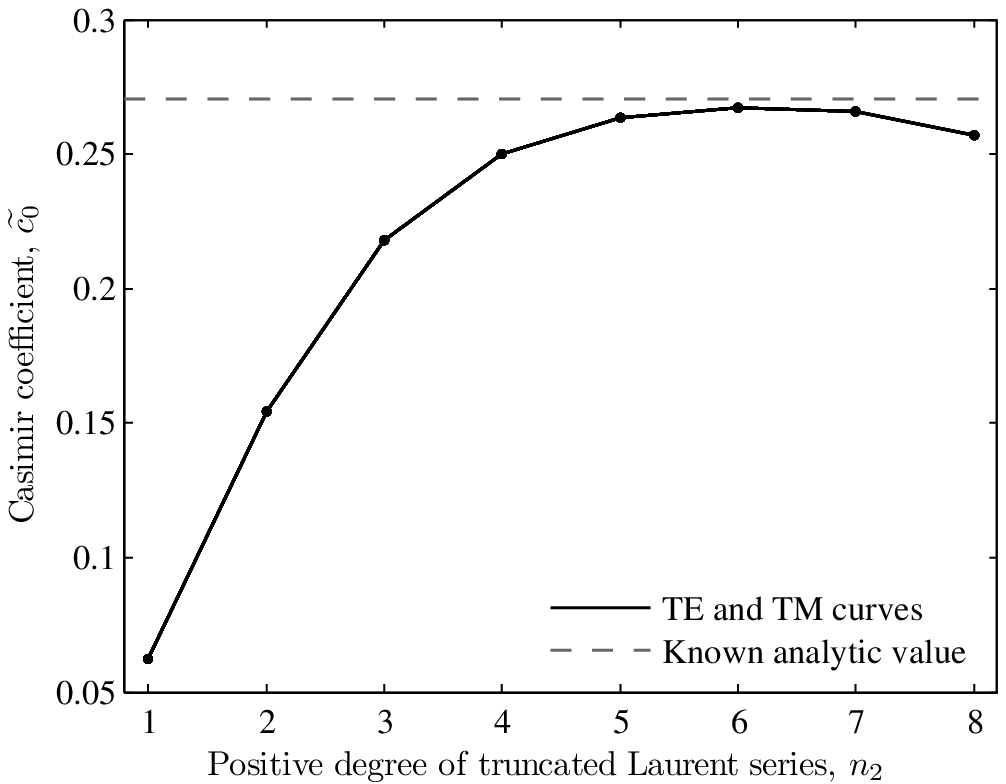}
    \caption{For data based on (\ref{data1}), the algorithm determines ${\cal N}=-4$ and a collection of $[x,y]$ points $[\wh{n}_{2},{\ccc_{0}}({\cal N}, \wh{n}_{2}, n_{2})]$ with $1 \le n_{2} \le 8$. The 8 curves obtained by joining these points are indistinguishable in this figure and the average of $\wt{c}_{0}({\cal N}, \wh{n}_{2}, n_{2})$ at their turning points yields a Casimir coefficient $c_{0}({\cal N})=0.27281$. The horizontal dotted line indicates the value of the Casimir coefficient that determines the observed Casimir attractive pressure between perfectly conducting plates separated by the vacuum.}
\end{figure}

\noindent The total (TE$+$TM) Casimir force per unit area on each face is attractive with magnitude:
\begin{eqnarray*}
    \left\vert\frac{1}{L_{x} L_{y}}\frac{\partial}{\partial L_{z}}\, \langle{ 2\cal E^{\TE}}\rangle \right\vert &=& \frac{\pi^{2}\hbar c}{240 L_{z}^{4}}
\end{eqnarray*}
and hence the force \textit{difference} between the plates is zero. Suppose now that the box is filled with an incompressible dielectric with inhomogeneous permittivity
\begin{eqnarray*}
    \epsilon(x,y,z) &=& \epsilon_{0} \, \exp({\alpha z}/{L_{z}}), \qquad\quad  0\le z\le L_{z}
\end{eqnarray*}
for some real inhomogeneity parameter $\alpha$. A quantization of the electromagnetic field in the box \cite{GTW} can be performed in a gauge where the vector potential ${\mathbf A}$ satisfies the condition $\nabla \cdot (\epsilon {\mathbf A})=0$. For $\sss\in\{\text{TE,TM}\}$, the regularized force differences   $\langle\Delta{\cal F}^{(\sss)}(\sigma)\rangle$, derived from the quantum expectation value of relevant components of the electromagnetic stress-energy-momentum tensor between the faces at $z=0$ and $z=L_{z}$ contributed by the TE and TM modes in the dielectric is, for $L_{x},\,L_{y} \gg L_{z}$, derived from a Laurent $s$-expansion of
\begin{eqnarray}\label{data2}
    \hspace{-0.6cm} \mbox{\small $\displaystyle -\frac{\Delta { \cal F }^{(\sss)} (\tt,\sigma)}{ {\cal F}_0}$} = \mbox{\small $\displaystyle\int_{0}^{\infty} \!\nu\, e^{-\tt \nu} \,d\nu  \int_{0}^{\infty} \!\Delta^{(\sss)}(i y,  \nu,\sigma)\,e^{-\tt y }\,dy$}.
\end{eqnarray}
Here $\sigma=e^{\alpha/2}$ is a positive inhomogeneity parameter, $ {\cal F}_{0}={\hbar c \alpha^{4} L_{x} L_{y} }/{64 \pi^{2}  L_{z}^{4}}$ and
\begin{eqnarray*}
    \Delta^{(\sss)}(z ,\nu,\sigma) &=& \partial_{z}\ln\left( F^{(\sss)}(z,\nu,\sigma )\right) \\
    F^{\TE}(z,\nu,\sigma) &=& J_{\nu}(z)\, Y_{\nu}(\sigma z) - J_{\nu}(\sigma z) \, Y_{\nu}(z) \\
    F^{\TM}(z,\nu,\sigma) &=& \wt{J}_{\mu}(z)\,\wt{Y}_{\mu}(\sigma z) - \wt{J}_{\mu}(\sigma z) \,\wt{Y}_{\mu}(z),
\end{eqnarray*}
where $\mu=\sqrt{\nu^{2}+1}$ and, for any Bessel function $\fz$, $\wt{\fz}(z)=z \fz^{\prime}(z) + \fz(z)$.  The above double integrals involve the integration of products of modified Bessel functions with respect to both order and argument and have resisted analytical evaluation. For $s>0$ they can however  be calculated numerically (on a laptop) for various $\sigma$  and fitted by regression to a truncated Laurent expansion in the vicinity of $s=0$. Our algorithm clearly establishes that the principal part of the truncated Laurent series corresponds to a pole of order $4$ (as is the case for $\alpha=0$ when $\epsilon$ is the permittivity of the vacuum). It also determines the Casimir coefficients and hence the regularized stress differences between the plate faces for both the TE and TM modes. For example, with $\sigma=8/27$ the TM modes contribute a force difference $3.54704\times 10^{-28}\,{ L_x L_y} /{L_z^4 }$ Newtons. This is 0.27282 times the value of the total Casimir force on either plate in the vacuum. Similarly the TE modes contribute a force difference $3.46159\times10^{-28}\,{ L_x L_y} /{L_z^4 }$ Newtons. This is 0.26625 times the value of the total Casimir force on either plate in the vacuum. The similarity of these mode contributions to each other is somewhat surprising given the differences in the structure of the TE and TM spectrum generators.
\begin{figure}
    \includegraphics[width=0.8\columnwidth]{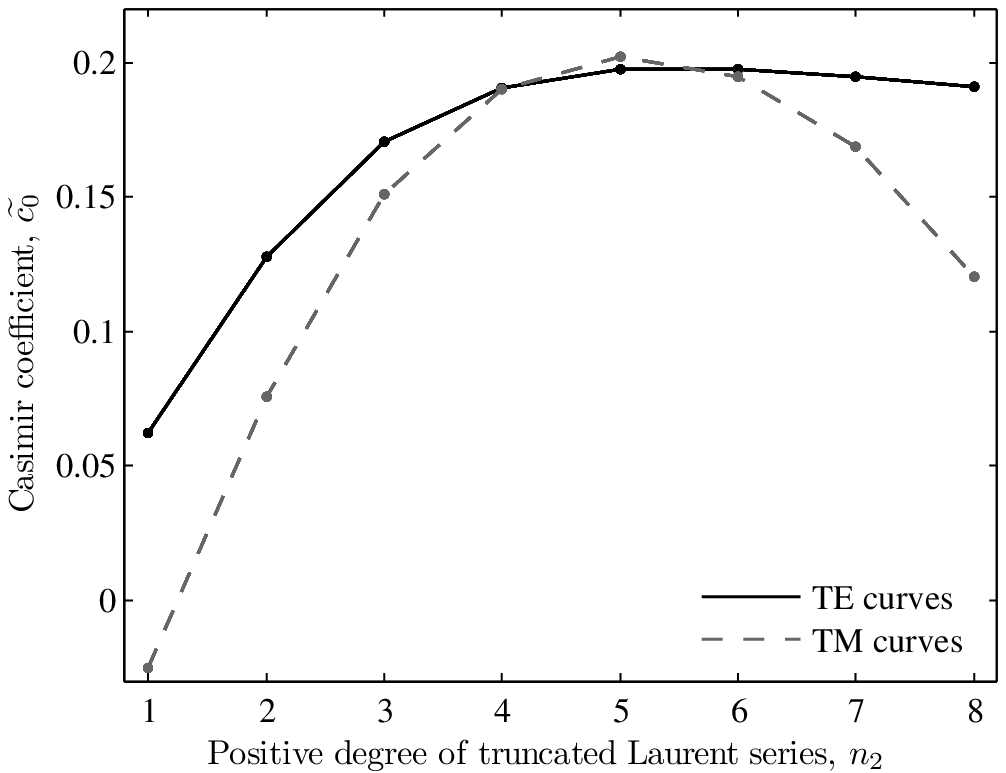}
    \caption{For data based on (\ref{data2}) with $\sigma={8}/{27}$ the algorithm determines ${\cal N}=-4$ and a collection of $[x,y]$ points $[\wh{n}_{2}  , {\ccc_{0}^{\TE}}({\cal N}, \wh{n}_{2}, n_{2})]$ and $[\wh{n}_{2},{\ccc_{0}^{\TM}}({\cal N}, \wh{n}_{2}, n_{2}) ]$  with $1 \le n_{2} \le 8$. The TE curves  obtained by joining the TE points are indistinguishable in this figure and the average of $\wt{c}_{0}^{\TE}({\cal N}, \wh{n}_{2}, n_{2})$  at their turning points yields a Casimir coefficient $c_{0}^{\TE}({\cal N})=0.19744$. The same is true for the TM points with  $c_{0}^{\TM}({\cal N})= 0.20231$. Each such  Casimir coefficient contributes to the total regularized force difference ${ \cal F }_{0}\left(c_{0}^{\TE}({\cal N}) + c_{0}^{\TM}({\cal N})\right) $,  between  opposite $z$-faces of a pair of perfectly conducting plates separated by an inhomogeneous dielectric with permittivity $\epsilon(x,y,z)=\epsilon_{0} \, \exp({\alpha z}/{L_{z}})$ where $\sigma=\exp({\alpha}/{2})$.}
\end{figure}

\section{Summary}
A robust numerical scheme for regularizing the quantum electromagnetic stresses in an inhomogeneous dielectric between conducting plates has been described. We believe that it has much wider applicability to more general systems such as those mentioned in the introduction.  Although the determination of electromagnetic  cavity modes in more complex geometries and media is in general non-trivial, once this hurdle is overcome the quantization program can proceed (with possible thermal corrections \cite{Lifshitz1,Philbin}) and the regularization method outlined here is then straightforward. However, any regularization scheme involving fields in media ultimately depends on the viability of the methods of field quantization in a non-dynamic background. It is therefore of paramount importance to verify precise results of such a scheme by experiment.

Any confined inhomogeneous material dielectric will sustain stresses induced by electromagnetic quantum fluctuations if the confining domain is rigid. If the medium remains static, such stresses induce mechanical (elastic) stresses in the dielectric to maintain equilibrium. Unlike similarly induced classical stresses by the classical gravitational field in the laboratory (that vary with the orientation of the dielectric), the quantum induced electromagnetic stresses are permanent. In principle, they could be detected experimentally by noting the variation of the induced stress field within the dielectric with variations of the permittivity inhomogeneities. Such variations might be detected using photo-elastic effects on the polarization of light passing through a transparent medium.


\vskip 1cm
    Goto and Walton are grateful for support from STFC and  the Cockcroft Institute and all authors would like to thank Ian Bailey and Steve Jameson for interesting discussions.
\bibliography{NumRegBib}

\begin{thebibliography}{20}%
\makeatletter
\providecommand \@ifxundefined [1]{%
 \@ifx{#1\undefined}
}%
\providecommand \@ifnum [1]{%
 \ifnum #1\expandafter \@firstoftwo
 \else \expandafter \@secondoftwo
 \fi
}%
\providecommand \@ifx [1]{%
 \ifx #1\expandafter \@firstoftwo
 \else \expandafter \@secondoftwo
 \fi
}%
\providecommand \natexlab [1]{#1}%
\providecommand \enquote  [1]{``#1''}%
\providecommand \bibnamefont  [1]{#1}%
\providecommand \bibfnamefont [1]{#1}%
\providecommand \citenamefont [1]{#1}%
\providecommand \href@noop [0]{\@secondoftwo}%
\providecommand \href [0]{\begingroup \@sanitize@url \@href}%
\providecommand \@href[1]{\@@startlink{#1}\@@href}%
\providecommand \@@href[1]{\endgroup#1\@@endlink}%
\providecommand \@sanitize@url [0]{\catcode `\\12\catcode `\$12\catcode
  `\&12\catcode `\#12\catcode `\^12\catcode `\_12\catcode `\%12\relax}%
\providecommand \@@startlink[1]{}%
\providecommand \@@endlink[0]{}%
\providecommand \url  [0]{\begingroup\@sanitize@url \@url }%
\providecommand \@url [1]{\endgroup\@href {#1}{\urlprefix }}%
\providecommand \urlprefix  [0]{URL }%
\providecommand \Eprint [0]{\href }%
\providecommand \doibase [0]{http://dx.doi.org/}%
\providecommand \selectlanguage [0]{\@gobble}%
\providecommand \bibinfo  [0]{\@secondoftwo}%
\providecommand \bibfield  [0]{\@secondoftwo}%
\providecommand \translation [1]{[#1]}%
\providecommand \BibitemOpen [0]{}%
\providecommand \bibitemStop [0]{}%
\providecommand \bibitemNoStop [0]{.\EOS\space}%
\providecommand \EOS [0]{\spacefactor3000\relax}%
\providecommand \BibitemShut  [1]{\csname bibitem#1\endcsname}%
\let\auto@bib@innerbib\@empty
\bibitem [{\citenamefont {Milton}\ \emph {et~al.}(1999)\citenamefont {Milton},
  \citenamefont {Nesterenko},\ and\ \citenamefont {Nesterenko}}]{Milton1}%
  \BibitemOpen
  \bibfield  {author} {\bibinfo {author} {\bibfnamefont {K.~A.}\ \bibnamefont
  {Milton}}, \bibinfo {author} {\bibfnamefont {A.~V.}\ \bibnamefont
  {Nesterenko}}, \ and\ \bibinfo {author} {\bibfnamefont {V.~V.}\ \bibnamefont
  {Nesterenko}},\ }\href@noop {} {\bibfield  {journal} {\bibinfo  {journal}
  {Phys. Rev. D}\ }\textbf {\bibinfo {volume} {59}},\ \bibinfo {pages}
  {105009:1} (\bibinfo {year} {1999})}\BibitemShut {NoStop}%
\bibitem [{\citenamefont {Milton}(1980)}]{Milton2}%
  \BibitemOpen
  \bibfield  {author} {\bibinfo {author} {\bibfnamefont {K.~A.}\ \bibnamefont
  {Milton}},\ }\href@noop {} {\bibfield  {journal} {\bibinfo  {journal} {Ann.
  Phys.}\ }\textbf {\bibinfo {volume} {127}},\ \bibinfo {pages} {49} (\bibinfo
  {year} {1980})}\BibitemShut {NoStop}%
\bibitem [{\citenamefont {Bordag}\ \emph {et~al.}(2009)\citenamefont {Bordag},
  \citenamefont {Klimchitskaya}, \citenamefont {Mohideen},\ and\ \citenamefont
  {Mostepanenko}}]{Bordag}%
  \BibitemOpen
  \bibfield  {author} {\bibinfo {author} {\bibfnamefont {M.}~\bibnamefont
  {Bordag}}, \bibinfo {author} {\bibfnamefont {G.~L.}\ \bibnamefont
  {Klimchitskaya}}, \bibinfo {author} {\bibfnamefont {U.}~\bibnamefont
  {Mohideen}}, \ and\ \bibinfo {author} {\bibfnamefont {V.~M.}\ \bibnamefont
  {Mostepanenko}},\ }\href@noop {} {\emph {\bibinfo {title} {{A}dvances in the
  {C}asimir {E}ffect}}}\ (\bibinfo  {publisher} {Oxford University Press,
  Oxford},\ \bibinfo {year} {2009})\BibitemShut {NoStop}%
\bibitem [{\citenamefont {Boyer}(1968)}]{Boyer}%
  \BibitemOpen
  \bibfield  {author} {\bibinfo {author} {\bibfnamefont {T.~H.}\ \bibnamefont
  {Boyer}},\ }\href@noop {} {\bibfield  {journal} {\bibinfo  {journal} {Phys.
  Rev.}\ }\textbf {\bibinfo {volume} {174}},\ \bibinfo {pages} {1764} (\bibinfo
  {year} {1968})}\BibitemShut {NoStop}%
\bibitem [{\citenamefont {Buhmann}\ and\ \citenamefont
  {Welsch}(2007)}]{Buhmann}%
  \BibitemOpen
  \bibfield  {author} {\bibinfo {author} {\bibfnamefont {S.~Y.}\ \bibnamefont
  {Buhmann}}\ and\ \bibinfo {author} {\bibfnamefont {D.-G.}\ \bibnamefont
  {Welsch}},\ }\href@noop {} {\bibfield  {journal} {\bibinfo  {journal} {Prog.
  Quant. Electr.}\ }\textbf {\bibinfo {volume} {31}},\ \bibinfo {pages} {51}
  (\bibinfo {year} {2007})}\BibitemShut {NoStop}%
\bibitem [{\citenamefont {Philbin}(2011{\natexlab{a}})}]{Philbin1}%
  \BibitemOpen
  \bibfield  {author} {\bibinfo {author} {\bibfnamefont {T.~G.}\ \bibnamefont
  {Philbin}},\ }\href@noop {} {\bibfield  {journal} {\bibinfo  {journal} {New
  J. Phys.}\ }\textbf {\bibinfo {volume} {12}},\ \bibinfo {pages} {123008}
  (\bibinfo {year} {2011}{\natexlab{a}})}\BibitemShut {NoStop}%
\bibitem [{\citenamefont {Philbin}\ \emph {et~al.}(2010)\citenamefont
  {Philbin}, \citenamefont {Xiong},\ and\ \citenamefont
  {Leonhardt}}]{Leonhardt1}%
  \BibitemOpen
  \bibfield  {author} {\bibinfo {author} {\bibfnamefont {T.~G.}\ \bibnamefont
  {Philbin}}, \bibinfo {author} {\bibfnamefont {C.}~\bibnamefont {Xiong}}, \
  and\ \bibinfo {author} {\bibfnamefont {U.}~\bibnamefont {Leonhardt}},\
  }\href@noop {} {\bibfield  {journal} {\bibinfo  {journal} {Ann. Phys.}\
  }\textbf {\bibinfo {volume} {325}},\ \bibinfo {pages} {579} (\bibinfo {year}
  {2010})}\BibitemShut {NoStop}%
\bibitem [{\citenamefont {Leonhardt}\ and\ \citenamefont
  {Simpson}(2011)}]{Leonhardt2}%
  \BibitemOpen
  \bibfield  {author} {\bibinfo {author} {\bibfnamefont {U.}~\bibnamefont
  {Leonhardt}}\ and\ \bibinfo {author} {\bibfnamefont {W.~M.~R.}\ \bibnamefont
  {Simpson}},\ }\href@noop {} {\bibfield  {journal} {\bibinfo  {journal} {Phys.
  Rev. D}\ }\textbf {\bibinfo {volume} {84}},\ \bibinfo {pages} {081701:1}
  (\bibinfo {year} {2011})}\BibitemShut {NoStop}%
\bibitem [{\citenamefont {Inui}(2003)}]{Inui1}%
  \BibitemOpen
  \bibfield  {author} {\bibinfo {author} {\bibfnamefont {N.}~\bibnamefont
  {Inui}},\ }\href@noop {} {\bibfield  {journal} {\bibinfo  {journal} {J. Phys.
  Soc. Japan}\ }\textbf {\bibinfo {volume} {72}},\ \bibinfo {pages} {280}
  (\bibinfo {year} {2003})}\BibitemShut {NoStop}%
\bibitem [{\citenamefont {Goto}\ \emph {et~al.}(2011)\citenamefont {Goto},
  \citenamefont {Tucker},\ and\ \citenamefont {Walton}}]{GTW}%
  \BibitemOpen
  \bibfield  {author} {\bibinfo {author} {\bibfnamefont {S.}~\bibnamefont
  {Goto}}, \bibinfo {author} {\bibfnamefont {R.~W.}\ \bibnamefont {Tucker}}, \
  and\ \bibinfo {author} {\bibfnamefont {T.~J.}\ \bibnamefont {Walton}},\
  }\href@noop {} {\bibfield  {journal} {\bibinfo  {journal} {Proc. SPIE}\
  }\textbf {\bibinfo {volume} {8072}},\ \bibinfo {pages} {80720O:1} (\bibinfo
  {year} {2011})}\BibitemShut {NoStop}%
\bibitem [{\citenamefont {Lifshitz}(1956{\natexlab{a}})}]{Lifshitz2}%
  \BibitemOpen
  \bibfield  {author} {\bibinfo {author} {\bibfnamefont {E.~M.}\ \bibnamefont
  {Lifshitz}},\ }\href@noop {} {\bibfield  {journal} {\bibinfo  {journal}
  {Soviet Phys. JETP}\ }\textbf {\bibinfo {volume} {2}},\ \bibinfo {pages} {73}
  (\bibinfo {year} {1956}{\natexlab{a}})}\BibitemShut {NoStop}%
\bibitem [{\citenamefont {Dzyaloshinskii}\ \emph {et~al.}(1961)\citenamefont
  {Dzyaloshinskii}, \citenamefont {Lifshitz},\ and\ \citenamefont
  {Pitaevskii}}]{Lifshitz3}%
  \BibitemOpen
  \bibfield  {author} {\bibinfo {author} {\bibfnamefont {I.}~\bibnamefont
  {Dzyaloshinskii}}, \bibinfo {author} {\bibfnamefont {E.}~\bibnamefont
  {Lifshitz}}, \ and\ \bibinfo {author} {\bibfnamefont {L.}~\bibnamefont
  {Pitaevskii}},\ }\href@noop {} {\bibfield  {journal} {\bibinfo  {journal}
  {Adv. Phys.}\ }\textbf {\bibinfo {volume} {10}},\ \bibinfo {pages} {165}
  (\bibinfo {year} {1961})}\BibitemShut {NoStop}%
\bibitem [{\citenamefont {{V}an Kampen}\ \emph {et~al.}(1968)\citenamefont
  {{V}an Kampen}, \citenamefont {{N}ijboer},\ and\ \citenamefont
  {{S}chram}}]{Kampen}%
  \BibitemOpen
  \bibfield  {author} {\bibinfo {author} {\bibfnamefont {N.~G.}\ \bibnamefont
  {{V}an Kampen}}, \bibinfo {author} {\bibfnamefont {B.~A.}\ \bibnamefont
  {{N}ijboer}}, \ and\ \bibinfo {author} {\bibfnamefont {K.}~\bibnamefont
  {{S}chram}},\ }\href@noop {} {\bibfield  {journal} {\bibinfo  {journal}
  {Phys. Lett. A}\ }\textbf {\bibinfo {volume} {26}},\ \bibinfo {pages} {307}
  (\bibinfo {year} {1968})}\BibitemShut {NoStop}%
\bibitem [{\citenamefont {Pendry}(1998)}]{Pendry}%
  \BibitemOpen
  \bibfield  {author} {\bibinfo {author} {\bibfnamefont {J.~B.}\ \bibnamefont
  {Pendry}},\ }\href@noop {} {\bibfield  {journal} {\bibinfo  {journal} {J.
  Mod. Opt.}\ }\textbf {\bibinfo {volume} {45}},\ \bibinfo {pages} {2389}
  (\bibinfo {year} {1998})}\BibitemShut {NoStop}%
\bibitem [{\citenamefont {Blau}\ \emph {et~al.}(1988)\citenamefont {Blau},
  \citenamefont {Visser},\ and\ \citenamefont {Wipf}}]{Visser}%
  \BibitemOpen
  \bibfield  {author} {\bibinfo {author} {\bibfnamefont {S.~K.}\ \bibnamefont
  {Blau}}, \bibinfo {author} {\bibfnamefont {M.}~\bibnamefont {Visser}}, \ and\
  \bibinfo {author} {\bibfnamefont {A.}~\bibnamefont {Wipf}},\ }\href@noop {}
  {\bibfield  {journal} {\bibinfo  {journal} {Nucl. Phys.B}\ }\textbf {\bibinfo
  {volume} {310}},\ \bibinfo {pages} {163} (\bibinfo {year}
  {1988})}\BibitemShut {NoStop}%
\bibitem [{\citenamefont {Ambj{\o}rn}\ and\ \citenamefont
  {Wolfram}(1983)}]{Wolfram}%
  \BibitemOpen
  \bibfield  {author} {\bibinfo {author} {\bibfnamefont {J.}~\bibnamefont
  {Ambj{\o}rn}}\ and\ \bibinfo {author} {\bibfnamefont {S.}~\bibnamefont
  {Wolfram}},\ }\href@noop {} {\bibfield  {journal} {\bibinfo  {journal} {Ann.
  Phys.}\ }\textbf {\bibinfo {volume} {147}},\ \bibinfo {pages} {33} (\bibinfo
  {year} {1983})}\BibitemShut {NoStop}%
\bibitem [{\citenamefont {Reuter}\ and\ \citenamefont
  {Dittrich}(1985)}]{Reuter}%
  \BibitemOpen
  \bibfield  {author} {\bibinfo {author} {\bibfnamefont {M.}~\bibnamefont
  {Reuter}}\ and\ \bibinfo {author} {\bibfnamefont {W.}~\bibnamefont
  {Dittrich}},\ }\href@noop {} {\bibfield  {journal} {\bibinfo  {journal} {Eur.
  J. Phys.}\ }\textbf {\bibinfo {volume} {6}},\ \bibinfo {pages} {33} (\bibinfo
  {year} {1985})}\BibitemShut {NoStop}%
\bibitem [{Note1()}]{Note1}%
  \BibitemOpen
  \bibinfo {note} {$\Psi (3,x) = \protect \frac {\partial ^{4}\protect \qopname
  \relax o{ln}( \Gamma (x)) }{\partial x^{4}}$}\BibitemShut {NoStop}%
\bibitem [{\citenamefont {Lifshitz}(1956{\natexlab{b}})}]{Lifshitz1}%
  \BibitemOpen
  \bibfield  {author} {\bibinfo {author} {\bibfnamefont {E.~M.}\ \bibnamefont
  {Lifshitz}},\ }\href@noop {} {\bibfield  {journal} {\bibinfo  {journal} {Zh.
  Eksp. Teor. Fiz.}\ }\textbf {\bibinfo {volume} {29}},\ \bibinfo {pages} {94}
  (\bibinfo {year} {1956}{\natexlab{b}})}\BibitemShut {NoStop}%
\bibitem [{\citenamefont {Philbin}(2011{\natexlab{b}})}]{Philbin}%
  \BibitemOpen
  \bibfield  {author} {\bibinfo {author} {\bibfnamefont {T.~G.}\ \bibnamefont
  {Philbin}},\ }\href@noop {} {\bibfield  {journal} {\bibinfo  {journal} {New
  J. Phys.}\ }\textbf {\bibinfo {volume} {13}},\ \bibinfo {pages} {063026}
  (\bibinfo {year} {2011}{\natexlab{b}})}\BibitemShut {NoStop}%
\end{thebibliography}%

\end{document}